\documentstyle[12pt,epsf]{article}
\begin{document}
\renewcommand{\theequation}{\thesection.\arabic{equation}}
\renewcommand{\section}[1]{\addtocounter{section}{1}
\vspace{5mm} \par \noindent
  {\bf \thesection . #1}\setcounter{subsection}{0}
  \par
   \vspace{2mm} } 
\newcommand{\sectionsub}[1]{\addtocounter{section}{1}
\vspace{5mm} \par \noindent
  {\bf \thesection . #1}\setcounter{subsection}{0}\par}
\renewcommand{\subsection}[1]{\addtocounter{subsection}{1}
\vspace{2.5mm}\par\noindent {\em \thesubsection . #1}\par
 \vspace{0.5mm} }
\renewcommand{\thebibliography}[1]{ {\vspace{5mm}\par \noindent{\bf
References}\par \vspace{2mm}}
\list
 {\arabic{enumi}.}{\settowidth\labelwidth{[#1]}\leftmargin\labelwidth
 \advance\leftmargin\labelsep\addtolength{\topsep}{-4em}
 \usecounter{enumi}}
 \def\newblock{\hskip .11em plus .33em minus .07em}
 \sloppy\clubpenalty4000\widowpenalty4000
 \sfcode`\.=1000\relax \setlength{\itemsep}{-0.4em} }
\newcommand\rf[1]{(\ref{#1})}
\def\nn{\nonumber}
\newcommand{\sect}[1]{\setcounter{equation}{0} \section{#1}}
\renewcommand{\theequation}{\thesection .\arabic{equation}}
\newcommand{\ft}[2]{{\textstyle\frac{#1}{#2}}}
\newcommand{\be}{\begin{equation}}
\newcommand{\ee}{\end{equation}}
\newcommand{\plb}[3]{{{\it Phys.~Lett.}~{\bf B#1} (#3) #2}}
\newcommand{\npb}[3]{{{\it Nucl.~Phys.}~{\bf B#1} (#3) #2}}
\newcommand{\prd}[3]{{{\it Phys.~Rev.}~{\bf D#1} (#3) #2}}
\newcommand{\ptp}[3]{{{\it Prog.~Theor.~Phys.}~{\bf #1} (#3) #2}}
\newcommand{\ijmpa}[3]{{{\it Int.~J.~Mod.~Phys.}~{\bf A#1} (#3) #2}}
\newcommand{\prl}[3]{{{\it Phys.~Rev.~Lett.}~{\bf #1} (#3) #2}}
\newcommand{\hepph}[1]{{\tt hep-ph/#1}}
\newcommand{\hepth}[1]{{\tt hep-th/#1}}
\newcommand{\leqsim}{\,\raisebox{-0.6ex}{$\buildrel < \over \sim$}\,}

\def\ep{\varepsilon}

\thispagestyle{empty}

\begin{center}

\vspace{3cm}

{\large\bf Some Thermodynamical Aspects of String Theory%
\footnote{From contributions by Eliezer
Rabinovici to 
``The Many Faces of the Superworld'': Yuri Golfand Memorial Volume, 
World Scientific (Singapore) 1999, and the EnglertFest, 
Universit\'e Libre de Bruxelles, 24-27 march 1999.
Based on a series of works with J.~Barbon, I.~Kogan
and S.~Abel \cite{br,bkr,abkr}.}}\\

\vspace{1.4cm}

{\sc S.A.~Abel$^{a}$, J.L.F.~Barb\'on$^{b}$, I.I.~Kogan$^{c}$ and
E.~Rabinovici$^{d}$}\\

\vspace{1.3cm}

{\em $a$ Service de Physique Th\'eorique, CEA-SACLAY, Gif-sur-Yvette, 
91191 France} \\
{\em $b$ Theory Division, CERN, CH-1211 Geneva 23, Switzerland} \\
{\em $c$ Theoretical Physics, 1 Keble Road, Oxford OX1 3NP, UK} \\
{\em $d$ Racah Institute of Physics, The Hebrew University, Jerusalem, 
Israel} \\

\vspace{1.2cm}

\end{center}

\baselineskip18pt
\addtocounter{section}{1}
\par \noindent
{\bf \thesection . Introduction}
  \par
   \vspace{2mm} 
\noindent

The possible phases of gauge theory have been uncovered and studied
even in the absence of an exact solution of such theories.
Their low energy physics can be classified according to the charges
of the dyons which condense (or not). In particular the Standard Model
utilizes three of the possible phases of gauge systems. 
The weak, colored and electromagnetic interactions correspond
to the condensation of electric, magnetic and no condensation
respectively.
A similar structure is yet to be uncovered in detail in theories
which contain gravity. Several pieces of information correlating
the phase structure of the worldsheet theory with that of the target
space theory are known. Theories which are in the topological phase
on the worldsheet lead to target space theories which are also topological
\cite{efr,baulieu}.
Theories which describe perturbatively strings moving in a background
of the form $(AdS)_{p+2} \times {\cal N}_{8-p}$ ($AdS$ stands for an
Anti-de Sitter spacetime and ${\cal N}$ for some appropriate compact manifold)
turn out to be well described by a $p+1$ dimensional target space theory 
\cite{baulieu} which is just a field
theory living on the boundary of the manifold \cite{mald}. 
There are quite a few
examples of this behavior. One is also familiar with the fact that 
perturbatively a string moving on a background of the form 
$R_{3,1} \times {\cal C}_6$ ($R_{3,1}$ is for example four dimensional 
Minkowski space and ${\cal C}_6$ is an appropriate compact manifold)
does not seem to be described by a regular field theory in target space
but rather by a theory with string like excitations which possesses
a very high degree of symmetry. 
There are indications that a system intermediate in some sense between
field theory and string theory may also exist \cite{seiberg}.
These are several  possible
phases of string theory. 
Here I will focus on some questions
raised by the field theoretical description of string theory on
$(AdS)_{p+2} \times {\cal N}_{8-p}$.


\setcounter{equation}{0}
\section{Should 4=10 be read in English or in Hebrew?}

\noindent 
Another title for this section could be `Extensivity vs. Holography'.
It had been suggested that in theories of gravity residing in $D$ 
spacetime dimensions the number of degrees of freedom (somehow 
suitably defined) should reflect a $D-1$ structure
\cite{holo}. In string theory
examples the reduction of degrees of freedom may seem even more
drastic, for example the propagation of a string on $AdS_5 \times
{\cal N}_5$ is described by an $N=4$ SUSY $SU(N)$ Yang-Mills theory
living in 4 spacetime dimensions. This relation can be approximately
described using a very smooth appropriate supergravity background when:
\begin{equation}
N \gg (g_{YM}^2 N)^{1/4} \gg 1.
\end{equation}
$N$ is the number of colors and $g_{YM}$ is the Yang-Mills gauge
coupling. 

We first reflect on the equation $4=10$ in Hebrew, that is we study
if it is possible 
and if it is true that what seems like a ten dimensional theory
actually exhibits four dimensional behavior. For this purpose the effective
spacetime dimensionality of the system is associated with the temperature
dependence of the entropy of the system at temperatures smaller than
string scale temperatures and larger than Kaluza-Klein temperatures.
For a temperature obeying:
\begin{equation}
m_{KK} \ll T \ll m_s, \qquad \qquad  VT^{D-1} \gg 1  \label{range}
\end{equation}
where $V$ is the spatial volume of the system, the entropy
is expected to behave as:
\begin{equation}
S \sim V T^{D-1} \label{entrfirst}
\end{equation}
from which one can read off the effective dimension $D$ of the system.
We will discuss two successive terms in the perturbation theory
around the supergravity background.
To leading order in $1/N^2$ common wisdom expects that 
the full quantum calculation of the entropy of the strongly coupled
gauge theory would be of the form
\begin{equation}
S=N^2 \,f(g_{YM}^2 \,N) \,V T^3 \label{entropy}
\end{equation}
where $f$ is an appropriate 
function of the Yang-Mills coupling. On the supergravity side
there are at least two classical backgrounds whose bulk geometry has
the same behavior on the four dimensional boundary on which the
Yang-Mills theory lives. One could imagine that one needs to sum over
all such bulk geometries which have the same boundary
\cite{hawk,witt}. It will turn
out that the failure to do so will not be consistent with the duality
conjecture. For the case at hand \cite{hawk,witt}, 
one such background which exists
for all values of the temperature is that of the $AdS_5$ at finite
temperature. The other background is that of a Schwarzschild-AdS black
hole. Both have a $S^3 \times S^1$ as a boundary. The temperature above
which the black hole is formed is proportional to the curvature energy
scale $1/b$,
of AdS space. The thermodynamical analysis on the supergravity side
indicates that the black hole configuration starts to dominate for temperatures
not much higher than that above which it may be formed. For those 
temperatures for which the AdS space dominates, the entropy vanishes
to order $N^2$. For high enough temperatures, the entropy is indeed
of the expected form (\ref{entropy}).
In this regime the classical supergravity calculation confirms holography
and reproduces the features of the 4 dimensional gauge theory.
The low temperature result is interpreted as reflecting a finite size
effect on the gauge theory side. For large $N$ the cooled YM theory
is thus supposed to pass 
to a phase in which the entropy is of order 1. An even more severe
test to the holography idea comes about to next order in $1/N$.
On the gauge theory side no qualitative changes of the formula
(\ref{entropy}) is expected. On the other hand on the supergravity 
side for temperatures (\ref{range}) one may expect the full ten
dimensionality of the system to be rediscovered.

Indeed, had the $AdS_5\times {\cal N}_5$ been the only contributing 
configuration, $D$ as appearing in (\ref{entrfirst}) would have been 9,
invalidating the holography property. More precisely, as shown in
Figure 1, the temperature in this background depends on a radial
coordinate $r$, while the KK gap $1/b$ is constant along $r$.

\begin{figure}[htp]
\begin{center}
\begin{picture}(130,240)
\setlength{\unitlength}{1mm}
\thicklines
\put(-10,55){\vector(0,1){30}}
\put(-10,55){\vector(1,0){65}}
\put(-17,80){\makebox(0,0){$T(r)$}}
\put(-13,64){\vector(0,1){10}}
\put(-13,64){\vector(0,-1){9}}
\put(-16,65){\makebox(0,0){$T_0$}}
\put(58,55){\makebox(0,0){\em r}}
\put(10,85){\makebox(0,0){$X_1$}}
\put(25,66){\makebox(0,0){$X_2$}}
\put(23,67){\vector(-1,-1){6}}
\put(-10,60){\line(1,0){65}}
\put(15,54){\line(0,1){10}}
\put(22,50){\makebox(0,0){$r_{crit}=b^2T_0$}}
\put(-4,57){\vector(0,1){3}}
\put(-4,57){\vector(0,-1){2}}
\put(1,57){\makebox(0,0){$1/b$}}
\put(-10,74){\line(3,-1){4}}
\put(-7,73){\line(2,-1){4}}
\put(-3,71){\line(3,-2){12}}
\put(9,63){\line(2,-1){8}}
\put(17,59){\line(3,-1){6}}
\put(23,57){\line(4,-1){4}}
\put(27,56){\line(1,0){5}}
%
\put(-10,5){\vector(0,1){30}}
\put(-10,5){\vector(1,0){65}}
\put(-17,25){\makebox(0,0){$S_T(r)$}}
\put(27,4){\line(0,1){2}}
\put(35,0){\makebox(0,0){$r_{crit}=b^2T_0$}}
\put(58,5){\makebox(0,0){\em r}}
\put(10,15){\makebox(0,0){$T^9$}}
\put(40,15){\makebox(0,0){$T^3$}}
\end{picture}
\end{center}
\caption{The radial dependence of the effective temperature
of the AdS manifold $X_1$ and the black hole $X_2$.
The bottom part shows the radial variation of the temperature
dependence of the entropy in the $X_1$. $1/b$ is the KK mass threshold.}
\label{Figure1}
\end{figure}
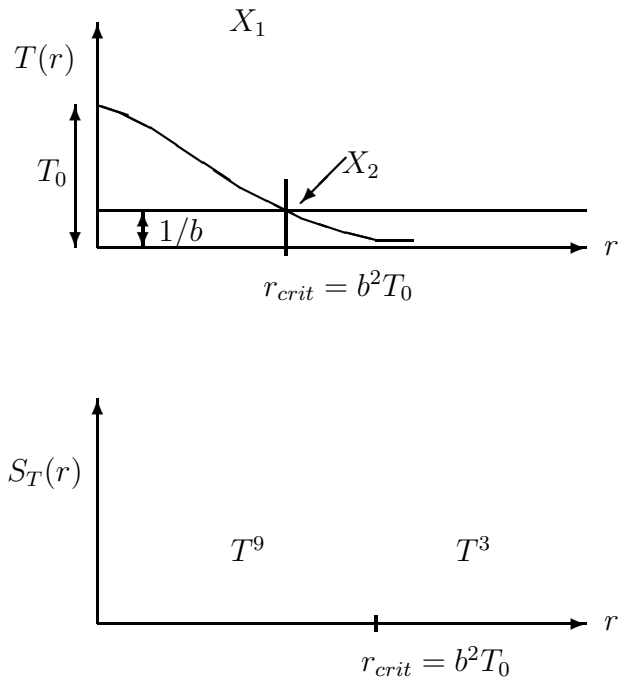

For large values of $r$, the temperature is red-shifted 
to very low values. For a temperature $T_0$ (the temperature at $r=0$)
larger than $1/b$ there
would be essentially two regions in the radial direction: for small $r$
the temperature would be hot enough to probe the full 10 dimensional structure
of the system; for $r$ above a certain critical radius the temperature
would be too cool to excite the KK modes and the naive effective dimension
of the system is 5. In fact, the red-shift in radial directions of AdS is
so strong that the true effective dimension
actually  drops to 4. The critical radius $r_c$ is proportional 
to the scale set by the AdS curvature $b$ and given by:
\begin{equation}
r_c=b^2 T_0.
\end{equation}
This apparent violation of holography is overturned by the emergence
of the second bulk supergravity configuration, namely that of the
black hole. Precisely for those temperatures at which holography
is at risk, the Euclidean black hole dominates the functional integral,
it hides beyond its horizon exactly that region in $r$ which
would have revealed the ten dimensional nature of the system. The
region in $r$ which actually exists for the black hole configuration
is cold, preserving the holographic four dimensional nature of 
the entropy. This is possible because the location of the
horizon of the black hole
is correlated to the radius of the compact manifold ${\cal N}$
(which dictates the KK transition temperature) in the appropriate
manner. Several lessons emerge: first it is essential to
sum over bulk geometries with different topologies in order to
enforce holography, and second the formation of black holes is
essential for the same purpose. 
One may also attempt to draw a lesson directly for string theory, that is,
given an initial perturbative background, non perturbative effects
in string theory would cause all backgrounds with the same boundary
(and perhaps other data) to contribute as well.
We have employed the methods 
used to derive the above results also for other string backgrounds
with non-constant negative curvature. In these cases, not only is the
temperature red-shifted for large values of $r$, but the KK mass
gap itself narrows for large values of $r$. These two effects are
competing. It turns out that Dirichlet $p$-branes continue  to
obey holography as long as $p$ is smaller than 5. $p=5$ is a marginal
case, and for $p>5$ an extensive dual field theory would not reproduce
the supergravity results. In this analysis it was important
that the black hole configuration dominated the AdS configuration.

As we turn to inspect the relation $4=10$ in English, more
respect will be paid to the `losing' configurations. The more precise
question is the following: suppose one could fully diagonalize the 
strongly coupled SUSY YM Hamiltonian on a finite spatial volume of
radius $R$. In that case  one could plot the density
of states as a function of the energy. Are there energy bands
for which the behavior of the density of states would be different
than that expected for a 4 dimensional theory? In particular would there
be a finite band for which the 4 dimensional system would exhibit
a 10 dimensional behavior? It was conjectured \cite{bdhm} 
that this is indeed the
case. The conjectured behavior is shown in Figure 2.

\begin{figure}[htp]
\begin{center}
\begin{picture}(130,200)
\setlength{\unitlength}{1mm}
\thicklines
\put(-30,5){\vector(0,1){70}}
\put(-30,5){\vector(1,0){100}}
\put(-37,65){\makebox(0,0){$S(E)$}}
\put(-25,4){\line(0,1){2}} 
\put(-20,0){\makebox(0,0){$E_{SUGRA}$}}
\put(-14,24){\makebox(0,0){$E^{9/10}$}}
\put(-25,10){\line(1,1){6}}
\put(-19,16){\line(3,2){6}}
\put(-13,20){\line(2,1){4}}
\put(-9,22){\line(4,1){4}}
\put(-5,23){\line(1,0){2}}
\put(-2,0){\makebox(0,0){$E_H$}}
\put(-3,4){\line(0,1){2}}  
\put(6,32){\makebox(0,0){$E$}}
\put(-3,23){\line(2,1){22}}
\put(17,0){\makebox(0,0){$E_{COR}$}}
\put(19,4){\line(0,1){2}} 
\put(32,37){\makebox(0,0){$E^{8/7}$}}
\put(19,34){\line(1,1){4}}
\put(23,38){\line(2,3){6}} 
\put(29,47){\line(1,2){4}}
\put(35,0){\makebox(0,0){$E_{Ads}$}}
\put(33,4){\line(0,1){2}} 
\put(46,58){\makebox(0,0){$E^{3/4}$}}
\put(33,55){\line(1,1){5}}
\put(38,60){\line(3,2){6}}
\put(44,64){\line(2,1){6}}
\put(50,67){\line(4,1){6}}
\put(73,5){\makebox(0,0){\em E}}
\end{picture}
\end{center}
\caption{The conjectured energy dependence of the entropy of 
$D=4$, maximally supersymmetric 
 Yang-Mills on a sphere of radius $R$. The following
definitions are used: $E_H \equiv \frac{1}{R} (g_s N)^{5/2}$,
$E_{COR} \equiv \frac{1}{R} N^2 (g_s N)^{-7/2}$,
$E_{Ads} \equiv \frac{1}{R} N^2$.}
\label{Figure2}
\end{figure}
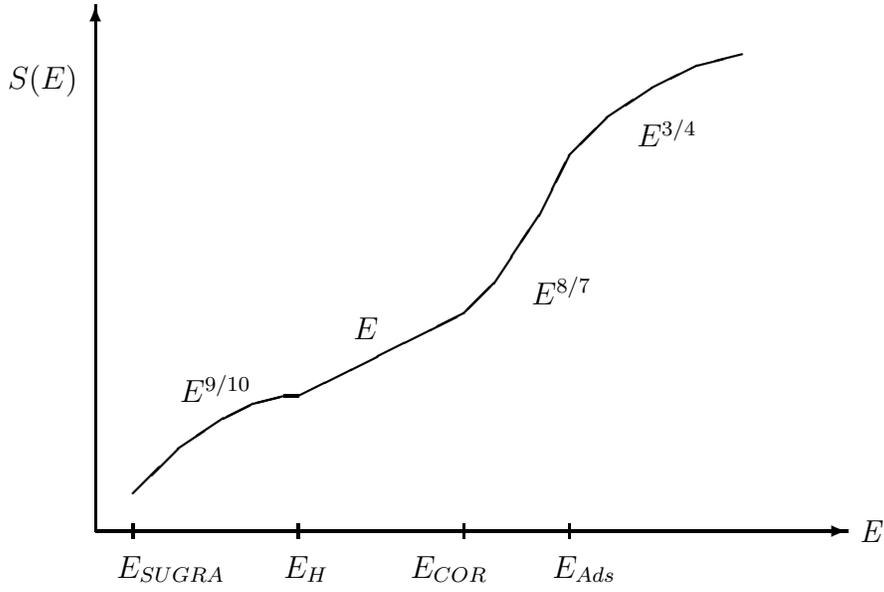

The curve  contains four sections: at the high energy end
the entropy $S(E)$ is proportional to $N^{1/2} \,(RE)^{3/4}$. 
This is the conventional
behavior of a 4 dimensional system with $N^2$ fields. 
However for other values of $E$
there are conjectured to be bands for which $S(E)$ is proportional
to $(RE)^{9/10}$ reflecting a ten dimensional 
behavior for a finite range of energy, bands for which $S(E) \sim
(g^2_{YM} N)^{-1/4} \,RE$ 
reflecting the usual perturbative Hagedorn spectrum of strings and finally
a finite size band for which  $S(E) \sim N^{-2/7} \,(RE)^{8/7}$ reflecting the
temporary formation of 10 dimensional Schwarzschild black holes.
Can one find any hint of this microcanonical behavior in the 
canonical analysis that we have performed on the supergravity side?
We believe that the answer is yes, and that the evidence can be
obtained by following the losing configuration. The 10 dimensional
behavior is hinted by the non-leading $AdS_5$ contribution to the
entropy which we have discussed earlier. Indeed, this contribution
comes from small values of $r$ which are short distance effects
in the supergravity
theory and thus long distance effects in the gauge theory exactly as
expected for finite size effects. The ten dimensional black holes
can be traced to black hole configuration with negative specific heat
which may form in AdS space above a certain temperature. These objects
will be localized or delocalized on ${\cal N}_5$ according
to the temperature. It is only the stringy Hagedorn regime which 
cannot be traced as easily on the supergravity side.
Actually some attempts to expose it have failed \cite{br,bkr} leading
to a conjecture of a `Hagedorn censorship'. However without 
a full stringy treatment this problem is still open. 
In conclusion
we have learned that indeed the relation $4=10$ should be read in
Hebrew, but as far as finite size effects are concerned, it is just
as interesting to view it in English. Imagine that we ourselves
are collecting data in some energy band misleading ourselves into
believing that the dimension of our space is larger than it really is...

\section{String thermodynamics in D-brane backgrounds}
The spectrum of hadrons in the dual model was given by the formula
\begin{equation}
S(E)= c E^a e^{\beta_H E}
\end{equation}
where $a$ and $\beta_H$ are determined by the theory.
This system has a limiting  temperature, the Hagedorn temperature $1/\beta_H$. 
After being amused by  encountering a
 limiting temperature, physicists suggested that the
limit was rather on our knowledge and that its emergence reflected the
existence of a phase transition. At temperatures around $1/\beta_H$
the system is much better described by underlying constituents 
of the hadrons, quarks and gluons \cite{cabi}. When the system is expressed in
terms of its constituents the temperature can be raised indefinitely
leading to the liberation of the confined quarks and gluons. Many string
theories have a similar density of states, the difference between 
the various theories is reflected by the values of the parameters
$a$ and $\beta_H$. Over the years the thermodynamics of open and closed,
bosonic, supersymmetric and heterotic string theories
was studied in some detail (see \cite{frau,engl} among others), 
sometimes in a hope to uncover the existence
of constituents of strings, the stuff the strings are made of.

In ten non-compact
dimensions, open strings indeed exhibit a limiting temperature.
Closed strings on the other hand do not exhibit such a behavior.
It is indeed tempting to consider the possibility that the Hagedorn
threshold may be crossed in that case. However 
microcanonical studies of the system were necessary due to the large
fluctuations exhibited near the Hagedorn temperature. In certain
circumstances a non-extensive behavior emerged driven by the formation
of a single very large string. It was conjectured that such a large
(effectively tensionless?)  string signaled a genuine phase transition.
The system also seemed to exhibit a negative specific heat.
On the other hand in many cases where the system was regularized by embedding it
in a finite volume, it turned out that one cannot surpass the Hagedorn
and eventually the energy pumped into the system was distributed more
evenly among the string modes, winding modes restored in some cases
a positive specific heat. We reexamined all these issues in the
presence of D-branes. The details appear in \cite{abkr}. 
The flavor of the results can be reproduced by simple random walk
arguments.

In addition to providing a nice physical
interpretation and checks of the calculations, this point of view
leads 
to some possible generalizations beyond toroidal backgrounds.

For example consider the single-string 
distribution function $\omega (\ep)$ for
closed strings in  $D$ large 
space-time dimensions. The energy $\ep$ of  the
string is proportional to the length of the random walk. The number
of walks with a fixed starting point 
 and a given length $\ep$ grows exponentially as $\exp
\,(\beta_c\,\ep)$.
 Since the walk must be closed, this overcounts by a
factor of the volume of the walk, which we shall denote by 
 $V({\rm walk}) = W$. Finally, there is a
factor of $V_{D-1}$ from the translational zero mode, and a factor of
$1/\ep$ because any point in the closed string can be a starting point.
The final result is
\be
\label{rw}
\omega(\ep)_{\rm closed} \sim V_{D-1}\cdot {1\over \ep} \cdot
{e^{\beta_c\,\ep} \over W}
.
\ee
Now, the volume of the walk is proportional to $\ep^{(D-1)/2}$ if it is
well-contained in the volume $(R\gg \sqrt{\ep})$, or roughly $V_{D-1}$
if it is space-filling $(R\ll \sqrt{\ep})$. One has the known result
\be
\omega (\ep)_{\rm closed}
 \sim V_{D-1} \,{e^{\beta_c \,\ep} \over \ep^{(D+1)/2}}
\ee
in $D$ effectively non-compact space-time dimensions, and 
\be
\omega (\ep)_{\rm closed} \sim {e^{\beta_c \,\ep} \over \ep} 
\ee
in an effectively compact space.     

We can generalize this analysis to open strings in the presence
of branes for a general $({\rm D}p,{\rm D}q)$
sector by a slight modification of the combinatorics. The leading
exponential degeneracy of a random walk of length $\ep$ with a fixed
starting point in say the D$p$-brane is the same as for closed
strings: $\exp (\beta_c\,\ep)$.
Fixing also the end-point at a {\it particular} point of the D$q$-brane
requires the factor $1/W$ to cancel the overcounting, just as in the
closed string case. Now, both end-points  move freely   
in the part of each brane occupied by the walk. This gives
a further degeneracy factor
\be
(W_{NN} \,W_{ND})\cdot (W_{NN} \,W_{DN})
\ee
from the positions of the end-points.   $N$ and $D$ refer to Neumann
and Dirichlet boundary conditions. 
Finally, the 
overall translation of the walk in the excluded NN volume gives a
factor
$V_{NN} / W_{NN}$. The final result is: 
\be 
\omega (\ep)_{\rm open} \sim {V_{NN}\over W_{NN}} \cdot
 W_{NN+ND} \cdot W_{NN+DN} \cdot {1\over W} \cdot \exp\,(\beta_c\,\ep) \sim  
 {V_{NN} \over W_{DD}} \;   
\exp\,(\beta_c\,\ep) 
.\ee 
Thus, we find that the density of states is only sensitive to the 
effective volume of the random walk in DD directions. If the walk 
is well-contained in DD directions $(R_{DD} \gg \sqrt{\ep})$,  
we find $W_{DD} \sim \ep^{d_{DD}/2} $ and  
\be
\omega (\ep)_{\rm open} \sim {V_{NN} \over  
\ep^{d_{DD} /2}} \;\exp\,(\beta_c\,\ep) 
.
\ee 
On the other hand, if  it is space-filling in DD directions $(R_{DD} \ll
\sqrt{\ep})$, the DD-volume of the walk is just $W_{DD} \sim 
V_{DD}$ and we find  
\be 
\omega(\ep)_{\rm open} \sim {V_{NN} \over V_{DD}}  
\;\exp\,(\beta_c \,\ep) 
.
\ee

The random walk picture gives a geometric rationale for the 
                similarity between non-compact closed-string
and open-string densities of states. It is related to the fact 
that the random walk must `close on itself' in some effective
co-dimension (the full space for closed strings and the DD space
for open strings). Canonically, open strings 
attached to D$p$-branes in infinite transverse space 
 for $p<5$ thus have the non-limiting 
characteristics of closed strings in ten dimensions.
These formulas are very useful to determine the canonical and 
microcanonical behavior of the system in various environments, depending
on the values of compactification moduli. 

I wish to
note here one speculative feature which emerges out of the analysis. It turns
out that in a system containing a collection of D-branes, D$p$-branes
with $p\geq 5$ attract energy from their neighboring branes. One
possible result of being an energy sink could be the melting of
these branes, leaving the arena free for D$p$-branes with $p<5$.
One can find various counter-arguments to this scenario in \cite{abkr}. 
Nevertheless, we find this Darwinistic concept worthy of further investigation.

\section{Phases in Gravity}
We end this contribution with an impressionistic discussion of bulk
and boundary phase diagrams in string
theory at moderately small coupling.
 I believe such diagrams will come to play as an important
a role as those drawn some years ago for gauge theories.

\subsection{Bulk Phase Diagram}

A supergravity gas in ten dimensions has entropy
\be
S(E)_{\rm sgr} \sim V^{1/10} \, E^{9/10}
\ee
and can be matched to a bulk black hole  with entropy ($\alpha' \sim \ell_s^2 
=1$ throughout this section)  
\be
S(E)_{\rm bh} \sim E \,(g_s^2 \,E)^{1/7}
.\ee
The coexistence
 line $S_{\rm sgr} \sim S_{\rm bh}$  gives a black hole in equilibrium with
radiation in a finite volume, with energy of order
\be
E({\rm sgr}\leftrightarrow {\rm bh}) \sim {1\over g_s^2} \,
 (g_s^2 \,V)^{7/17}
,\ee
and  microcanonical temperature
\be
\label{tha}
T({\rm sgr}\leftrightarrow {\rm bh}) \sim \left({1\over g_s^2
\,V}\right)^{1/17}
.\ee
Since the black-hole-dominated region has negative specific heat, this
temperature is maximal in the vicinity of the transition.
 This configuration is microcanonically stable in finite
volume, in a range
 of energies between the matching point and the Jeans bound.

\begin{figure}[htp]
\begin{center}
\begin{picture}(130,240)
\setlength{\unitlength}{1mm}
\thicklines
\put(-30,5){\line(0,1){96}}
\put(-30,5){\line(1,0){91}}
\put(61,5){\line(0,1){96}}
\put(-30,101){\line(1,0){91}}
\put(-37,90){\makebox(0,0){$g_S$}}
\put(55,-2){\makebox(0,0){\em E}}
\put(24,100){\line(1,-4){10}}
\put(34,60){\line(1,-3){8}}
\put(42,36){\line(1,-2){6}}
\put(48,24){\line(2,-3){13}}
%
\put(-7,100){\line(1,-4){10}}
\put(3,60){\line(1,-3){8}}
\put(11,36){\line(1,-1){12}}
\put(23,24){\line(3,-2){9}}
\put(32,18){\line(2,-1){20}}
\put(52,8){\line(3,-1){9}}
%
\put(11,5){\line(0,1){31}}
\put(-10,40){\makebox(0,0){Sugra}}
\put(12,65){\makebox(0,0){Black}}
\put(15,60){\makebox(0,0){Hole}}
\put(43,80){\makebox(0,0){Holographic}}
\put(46,75){\makebox(0,0){Bound}}
\put(19,15){\makebox(0,0){Hag}}
\end{picture}
\end{center}
\caption{An impressionistic bulk phase diagram. Only the region
$g_s<1$ is represented in this picture. The triple point separating
the supergravity gas, black hole, and Hagedorn-dominated regimes is
located at $g_s \sim 1/\sqrt{V}$, and $E \sim V$. The rightmost region
is excluded by the holographic bound.}
\label{Figure3}
\end{figure}
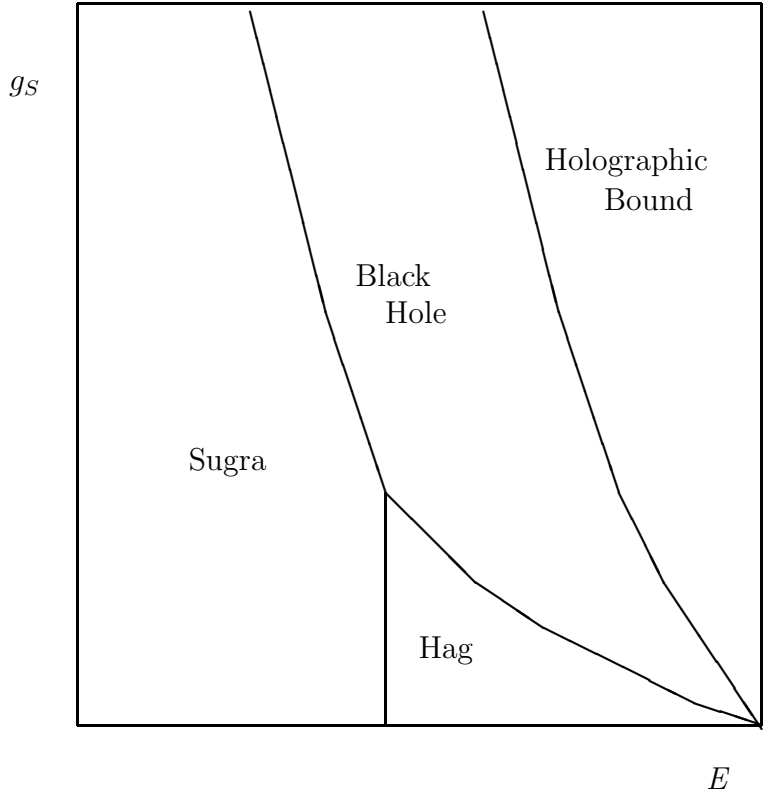

 The graviton
gas can also  be matched
  to a gas of long closed strings. The coexistence curve $S_{\rm sgr} \sim
S_{\rm Hag}$ at temperatures $T\sim O(1)$ in string units,
  is independent
of the string coupling and is given by the Hagedorn energy density:
\be
E({\rm sgr}\leftrightarrow {\rm Hag}) \sim V
.\ee
  This Hagedorn phase can be exited at high energy or large
coupling through the
 correspondence curve $S_{\rm Hag} \sim S_{\rm bh}$:
\be
E({\rm bh}\leftrightarrow {\rm Hag}) \sim {1\over g_s^2} 
,\ee
into a black-hole dominated phase at lower temperatures. 
The resulting
phase diagram for
the bulk or closed-string sector is depicted in Figure \ref{Figure3}.

\vspace{1cm}

An interesting feature of the phase diagram 
 is the existence of a triple point at the intersection of
 the phase boundaries of
the massless supergravity gas, Hagedorn,
 and black-hole-dominated regimes.  This point lies at Hagedorn
energy density
 $E_c \sim V$, string scale temperatures $T\sim O(1)$,
 and considerably weak coupling $g_s \sim 1/\sqrt{V}$ and, somewhat 
optimistically, we would like to interpret its
existence 
 as evidence for completeness of this phase structure. Namely, we are not
missing any major
 set of degrees of freedom. According to this picture, the Hagedorn phase
goes into a 
 black-hole-dominated phase at large energy or coupling, well within the
Jeans or holographic bound:
\be
\label{holbulk}
E<E_{\rm Hol} \sim  {V^{7/9} \over g_s^2}
,\ee
provided we are at
 weak string coupling $g_s <1$.  We see that the Hagedorn regime has no
thermodynamic limit
 whatsoever. If we scale the total energy $E$ linearly with the volume,
we run into the black-hole phase, which ends when   the horizon crushes
the walls of the box (i.e. the black hole fills the box).
 Moreover, if the string coupling is larger that $1/\sqrt{V}$, we miss
the Hagedorn phase 
altogether, as the supergravity gas goes into 
the black-hole-dominated phase directly.
In this case, the system has a sub-stringy maximum temperature
\be\label{tma}
T_{\rm max} \sim T({\rm sgr}\leftrightarrow {\rm bh}) < 1.
\ee

\subsection{World-Volume Phase Diagram}

\begin{figure}
\epsfysize=2.5 true in
\hskip 70 true pt
\epsfbox{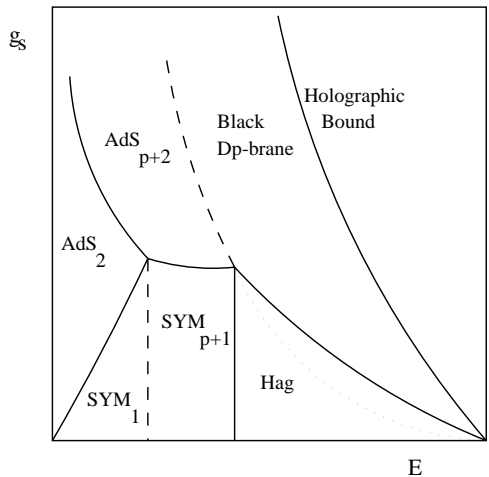}
\caption{
An impressionistic world-volume phase diagram for small string
coupling.
}
\label{Figure4}
\end{figure}

Similar remarks apply to the
 open-string sector in the vicinity of the D-branes.
In Figure \ref{Figure4} the world-volume phase diagram is
presented for small string coupling.
Here, the details of the correspondence principle depend on the
excitation energy of the D-brane, i.e. in the geometric picture,
we must distinguish between the near-extremal $(r_0 \ll r_Q)$ and
non-extremal or Schwarzschild $(r_0 \gg r_Q)$ regimes.    

Before proceeding further, it 
 is important to notice that D$p$-branes with $p>6$
cannot be considered as well-defined asymptotic states in weakly-coupled
string theory. The massless fields specifying the closed-string vacuum,
including the dilaton,
grow with transverse distance to the D-brane. As a consequence,
introducing a $p>6$ D$p$-brane in a given perturbative background
inevitably results in a non-perturbative modification of the vacuum
itself. Thus, consistency with the requirement of weak string coupling
throughout the system means that such branes are never far from
orientifold boundaries, and should be better considered as part of the
specification of the background geometry.
In the following, we shall restrict to $p<7$, unless specified otherwise.

  The matching
of the near-extremal $(r_0 \ll r_Q)$ 
black-brane entropy or Anti- de Sitter-type
 (AdS) throats:
\be
\label{sads}
S(E)_{{\rm AdS}_{p+2}} 
\sim N^{1/2} \, 
(V_{\parallel})^{5-p
 \over 2(7-p)} \, g_s^{p-3 \over 2(7-p)} \, E^{9-p \over 
2(7-p)} 
\ee
to a weakly-coupled Yang--Mills gas on the world-volume:
\be
\label{sym}
S(E)_{{\rm SYM}_{p+1}}
 \sim N^{2\over p+1} \, (V_{\parallel})^{1\over p+1} \, E^{p\over p+1}
,\ee
 is the content of the
 generalized SYM/AdS
 correspondence~\cite{mald}, and was studied in detail in
\cite{maldai,bkr,mart} (we call these manifolds AdS although, properly
speaking,  they
are only conformal to $AdS_{p+2} \times {\bf S}^{8-p}$).

 There are interesting finite-size effects at low temperatures,
$T \leqsim 1/R_{\parallel}$,  in the form of
large $N$ phase transitions of the gauge theory. 
For $p=3$ and spherical topology of the brane world-volume, the
gravitational counterpart is the Hawking--Page transition \cite{hawk,witt}
between the AdS black-hole geometry and the AdS vacuum geometry
(intermediate metastable phases can be found \cite{bdhm}). For
our case ($p<7$ and toroidal topology of the branes) the finite-size
effects setting in at the energy threshold $E\leqsim N^2 /R_{\parallel}$ 
are associated to the transition to
 zero-mode dynamics in the Yang--Mills language and  
  to finite-volume  localization
 \cite{grego} in the black-hole language.
At sufficiently low temperatures one must use a T-dual description of
the throat, resulting in an effective geometry of `smeared' D0-branes.
When these D0-branes localize as in \cite{grego} the description
involves an AdS-type throat with $p=0$, which we denote by ${\rm
AdS}_{2}$. In this case of toroidal topology,
 there is no regime of {\it vacuum} AdS dominance,
provided $N$ is large enough \cite{bkr,pr}.
 We refer the reader to \cite{bkr,mart,pr}
for
a detailed discussion of such low-temperature phenomena, as well as for
an extension of the phase diagram to large values of the string 
coupling, beyond
the 't Hooft limit discussed here.

  At temperatures $T>1/R_{\parallel}$ these finite-size effects can be
neglected, and the SYM/AdS transition is 
determined by the matching of (\ref{sads}) and (\ref{sym}).
 The transition temperature,
\be
\label{adst}
T({\rm SYM}_{p+1}\leftrightarrow {\rm AdS}_{p+2})
 \sim (g_s\,N)^{1\over 3-p}
,\ee
is smaller than the Hagedorn temperature as long as stringy energy
densities are not reached in the world-volume.

 At this point, it
should be noted that the interpretation of the AdS throats as SYM
dynamics at large 't Hooft coupling (the standard AdS/SYM
correspondence) is problematic for $p=5,6$. For $p=5$, the AdS regime
has a density of states typical of a string theory, with renormalized
tension $T_{\rm eff} = 1/\alpha' \,g_s\,N$.
 For $p=6$ the
qualitative features of the 
thermodynamics of the near-extremal and Schwarz\-schild regimes are  
essentially the same, so that the boundary $r_0 \sim r_Q$ does not
mark a significant change in behaviour. The holography
properties  required 
to interpret the AdS physics {\it only} in terms of gauge-theory dynamics  
 seem to break down for these cases \cite{malstro,br,kutse,bkr,polpeet}.
  However, the SYM/AdS
correspondence line in the sense of \cite{polhor} can always be
defined, independently of whether there is a candidate 
 microscopic interpretation 
for the entropy (\ref{sads}) in the AdS regime.                

At stringy energy densities $E\sim N^2 \,V_\parallel$, the SYM/AdS
correspondence line joins the open-string Hagedorn regime.  
The transition from a Yang--Mills gas on the world-volume
to a Hagedorn regime of  open strings $(S_{\rm SYM} \sim S_{\rm Hag})$
occurs at the energy 
\be
E({\rm SYM}_{p+1}\leftrightarrow {\rm Hag}) \sim N^2 \, V_{\parallel}
.\ee
This line joins the 
SYM/AdS correspondence curve at a {\it triple} point 
(see Figure \ref{Figure4}), the other phase boundary being the correspondence
curve between the long open strings in the Hagedorn phase, and the
non-extremal black-brane phase.  
Black D$p$-branes in the  Schwarzschild regime
$(r_0 \gg r_Q)$ have  entropy:
\be
S(E)_{{\rm B}p} \sim E\, 
\left({g_s^2 E \over V_{\parallel}}\right)^{1\over 7-p}
.\ee
and match the world-volume Hagedorn phase along the curve:
\be
\label{tr}
E({\rm Hag}\leftrightarrow {\rm B}p) \sim {V_{\parallel} \over g_s^2}
.\ee

Notice that the boundary line    
separating the near-extremal (AdS) and Schwarzschild (B$p$) regimes
of the black branes, given by
$r_0 \sim r_Q$, or   
\be
\label{satrad}
E({\rm AdS}_{p+2}
\leftrightarrow {\rm B}p) \sim N\;{V_{\parallel} \over g_s}
,\ee
also joins the triple point located at $E\sim N^2 \,V_\parallel$ and $g_s
\,N\sim 1$.   The temperature along this line is
\be
\label{tb}
T({\rm AdS}_{p+2} \leftrightarrow {\rm B}p) \sim \left({1\over g_s
\,N}\right)^{1\over 7-p}
.\ee
This temperature is locally maximal for small energy variations
if $p<5$.  

All these phases lie well
 within the holographic bound, defined by the condition that the
 horizon of the
black brane saturates the available transverse volume:
\be
\label{holwv}
E<E_{\rm Hol} \sim 
{V_{\parallel} \over g_s^2} \cdot (V_{\perp})^{7-p \over 9-p}
.\ee

In all of the above, supersymmetry was the
{\em \'eminence grise}. In its absence none of the above calculations
could have been consistently done.

\setcounter{equation}{0}

\vspace{.75cm}


\vspace{5mm}





\begin{thebibliography}{99}
\bibitem{br} J. L. F. Barbon, E. Rabinovici,
Extensivity Versus Holography in Anti-de Sitter Spaces,
Nucl.Phys. B545 (1999) 371; \hepth{9805143.}
\bibitem{bkr} J. L. F. Barbon, I. I. Kogan, E. Rabinovici,
On Stringy Thresholds in SYM/AdS Thermodynamics,
Nucl.Phys. B544 (1999) 104; \hepth{9809033.} 
\bibitem{abkr} S. A. Abel, J. L. F. Barbon, I. I. Kogan, E. Rabinovici,
String Thermodynamics in D-Brane Backgrounds,
JHEP 9904 (1999) 015; \hepth{9902058.} 
\bibitem{efr} S. Elitzur, A. Forge and E. Rabinovici, {\it Nucl. Phys.}
{\bf B388} (1992) 131.
\bibitem{baulieu} L. Baulieu and E. Rabinovici, work in progress.
\bibitem{mald} J.M. Maldacena, {\it Adv. Theor. Math. Phys.} {\bf 2} (1998)
231, \hepth{9711200;} S.S. Gubser, I.R. Klebanov and A.M. Polyakov,
{\it Phys. Lett.} {\bf B428} (1998) 105, \hepth{9802109;} E. Witten,
{\it Adv. Theor. Math. Phys.} {\bf 2} (1998) 253, \hepth{9802150;} 
O. Aharony, S.S. Gubser, J.M. Maldacena, H. Ooguri and Y. Oz, \hepth{9905111.}
\bibitem{seiberg} N. Seiberg, {\it Phys. Lett.} {\bf 408B} (1997) 98,
\hepth{9705221.} 
\bibitem{holo} G. 't Hooft, {\tt gr-qc/9310026;} L. Susskind, {\it J. Math.
Phys.} {\bf 36} (1995) 6377, \hepth{9409089.} 
\bibitem{hawk} S.W. Hawking and D. Page, {\it Comm. Math. Phys.} {\bf 78B}
(1983) 577.
\bibitem{witt} E. Witten, {\it Adv. Theor. Math. Phys.} {\bf 2} (1998)
505, \hepth{9803131.} 
\bibitem{bdhm} T. Banks, M.R. Douglas, G.T. Horowitz and E. Martinec,
\hepth{9808016.} 
\bibitem{cabi} N. Cabibbo and G. Parisi, {\it Phys. Lett.} {\bf B59} (1975) 67.
\bibitem{frau} S. Frautschi, {\it Phys. Rev.} {\bf D3} (1971) 2821;
R. Carlitz, {\it Phys. Rev.} {\bf D5} (1972) 3231.
\bibitem{engl} F. Englert and J. Orloff, {\it Nucl. Phys.} {\bf B334}
(1990) 472.

\bibitem{maldai}
N. Itzhaki, J.M. Maldacena, J. Sonnenschein and S. Yankielowicz,
\prd{58}{046004,}{1998} \hepth{9802042.}
\bibitem{mart}
M. Li, E. Martinec and V. Sahakian, \hepth{9809061};
E. Martinec and
V. Sahakian, \hepth{9810224;} \hepth{9901135.}

\bibitem{grego}
R. Gregory and R. Laflamme, \prl{70}{2837,}{1993} \hepth{9301052.}
\bibitem{pr}
A.W. Peet and S.F. Ross, \hepth{9810200.}
\bibitem{malstro}
J.M. Maldacena and A. Strominger,
  {\it J. High Energy Phys.} {\bf 12} (1997) 008,
 \hepth{9710014.}
\bibitem{kutse}
O. Aharony, M. Berkooz, D. Kutasov and N. Seiberg, \hepth{9808149.}
\bibitem{polpeet}
A.W. Peet and J. Polchinski, \hepth{9809022.}
\bibitem{polhor}
G.T. Horowitz and J. Polchinski, \prd{55}{6189,}{1997} \hepth{9612146.}



\end{thebibliography}
\end{document}